\documentclass[conference]{IEEEtran}
\IEEEoverridecommandlockouts

\usepackage[T1]{fontenc}
\usepackage{amsmath,amssymb,amsfonts}
\usepackage{algorithm}
\usepackage{algpseudocodex}
\usepackage{graphicx}
\usepackage{textcomp}
\usepackage[caption=false]{subfig}
\usepackage{float}
\usepackage{xcolor}
\usepackage{booktabs, multirow}
\usepackage{siunitx}
\usepackage{xspace}

\usepackage[colorlinks=true,allcolors=black]{hyperref}
\usepackage[numbers,sort&compress]{natbib}
\ExplSyntaxOn

\ExplSyntaxOff

\title{Profiling the Effective Limits of Error Mitigation via Circuit Replication}
\author{%
  \IEEEauthorblockN{Jeremie Pope}
  \IEEEauthorblockA{\textit{School of Computer Science and Engineering} \\
    \textit{Pennsylvania State University}\\
    Centre County, PA, USA \\
    j.pope@psu.edu}
  \and
  \IEEEauthorblockN{Swaroop Ghosh}
  \IEEEauthorblockA{\textit{School of Computer Science and Engineering} \\
    \textit{Pennsylvania State University}\\
    Centre County, PA, USA \\
    szg212@psu.edu}
}

\begin{document}
\maketitle

\begin{abstract}
  Current era quantum computers continue to grow in both capability and capacity. Despite these advancements, errors induced by environmental noise severely limit practical applicability. Current research into error mitigation and correction to bridge the gap between current-era quantum computers and the execution of noise-sensitive workloads. These methods have significant performance and resource overheads, thereby greatly limiting the real-world benefits of their use. Circuit replication, as a naive form of error mitigation, is not new and has largely been ignored given the resource constraints of current quantum hardware. However, its simplicity is attractive as a means to supplement modern methods, reducing the overall performance overhead while still preserving error-mitigation capabilities. In this paper, we profile the effects of simple circuit replication under real-world noise profiles to better establish replication's limits as a supplemental mitigation strategy.
  Quantum Approximate Optimization Algorithm (QAOA) for the Maxcut problem is explored for the analysis. For small graphs, we found that the average inference strength decreases by approximately $21.8\%$ while the average standard deviation decreases by $108.8\%$ compared to $6$ replicates. For larger graphs, inference strength decreases by $35.4\%$ while the average standard deviation decreased only $20.5\%$. Fewer replications did not affect smaller graphs, but degraded inference strength, with comparable benefits to standard deviation in larger graphs. These results show that replication has potential uses as a supplemental mitigation strategy for large-depth, highly variable workloads.
\end{abstract}


\section{Introduction}

Current-era quantum computers still suffer from limited resources and high error rates \cite{2020/Leymann,2025/Thakur}, severely limiting their usefulness for problems that are unsolvable by classical means. While advancements in materials science and manufacturing have increased the number of raw, high-quality qubits available for programmers to use, the industry at large is still far from being able to implement the logical qubits that would otherwise allow for fault-tolerant quantum computations at scale \cite{2024/Kahn,2025/Thakur,2025/Zhang}. In the absence of hardware solutions, quantum programs running on current-era Noisy Intermediate-Scale Quantum (NISQ) computers leverage quantum error mitigation and correction techniques to increase output quality. 
However, these techniques place greater constraints on available compute resources, particularly in the number of raw qubits required for execution \cite{2025/Thakur}. Programmers using error correction techniques must carefully balance the level of correction against the added cost of additional computing resources. Of particular concern is the depth of a quantum program: the more gates it contains, the greater the chance that environmental noise, decoherence, or other sources of error accumulate and ruin the final calculation. While error correction can fix some of this error, they also greatly increase circuit depth, thereby introducing more noise into the final output. A natural question that arises from this situation is: at what point is one method more favorable than another?

Establishing this point requires benchmarking which is incredibly difficult because there are no widely accepted metrics for comparing algorithms \cite{2024/Nath}. Worse still, the hardware platform, workload, or transpiler configuration affects collected metrics, muddying comparative analysis between approaches. As a result of these combinatorial complications, the examination of quantum algorithms has largely focused on the behavior of individual methods as applied to individual workloads. The use of multiple methods in tandem has largely gone unconsidered, with prior works focusing on smaller-scoped corrections via Pauli twirling \cite{2016/Wallman}, dynamical decoupling \cite{2023/Ezzel}, or noise extrapolation \cite{2020/Giurgica-Tiron,2023/Berg}, among others. We found no works to date that have examined whether higher-performing, complex methods can leverage lower-performing, simpler methods to reduce resource costs. To that end, we are investigating whether using naive circuit replication can reduce the error which in turn can be leveraged to simplify error-corrected quantum programs while still preserving some, if not all, of the original output quality.

Circuit replication as a form of error mitigation is neither new nor novel \cite{2019/Tannu,2024/Djidjev,2024/Parry}. In the strictest sense, it is not a form of error mitigation at all; instead, it relies on multiple programs executed in parallel to accumulate errors at different rates. If the error rates between replicated programs are sufficiently different, the combination of outputs leads to the desired answer having a higher probability than erroneous answers on average. This technique trades increased depth for increased width, potentially reducing the need/overhead of more costly and complex methods such as quantum error correction. To begin this analysis, the performance characteristics of circuit replication in relation to error mitigation need to be well understood.

In this paper, we present our initial findings on how circuit replication behaves under controlled noise models. Using Qiskit \cite{2024/Qiskit}, a common platform for executing quantum applications, we show how different levels of replication influence the quality of output for the Maxcut problem using a Quantum Approximate Optimization Algorithm (QAOA) on IBM's Heron R2 hardware. We show that $6$ replicates can reduce overall variability in output by up to $108.8\%$ but at the cost of $21.8\%$ worse inference strength versus no replication. We also show that as few as $3$ replicates achieve comparable reductions in overall variability, with slightly less impact on overall inference strength. 

The rest of the paper is structured as follows: 
Section~\ref{sec:background} provides further detail on prior applications of circuit replication.
Section~\ref{sec:methodology} outlines our analysis methodology and explains the workload chosen.
Section~\ref{sec:evaluation} presents the experimental results and provides the interim analysis.
Finally, we summarize key findings and provide a final analysis in Section~\ref{sec:conclusion}.

\section{Background}
\label{sec:background}

In this section, we describe how circuit replication is used as a error mitigation strategy while also detailing some of the inherent restrictions of the method. We also provide background on the chosen workload used as the comparative basis for this analysis.

\subsection{Utilizing Circuit Replication}

In quantum program replication
\cite{2019/Tannu,2024/Djidjev,2024/Parry}, a programmer can replicate a single computation in parallel to produce a composite output that, when combined, has a higher average probability of producing the correct answer on average in a noisy environment. This approach works so long as each replicated computation operates under dissimilar error rates. If computations share a similar error domain, erroneous solutions across replicated computations become correlated, inadvertently amplifying suboptimal solutions in the final average \cite{2019/Tannu,2024/Djidjev}.

For quantum computers, requiring differing error rates between replicates poses a problem as NISQ computers are greatly influenced by the environmental noise intrinsic to the entire itself, impacting all qubits within the machine in a correlated way. In adapting replication for this domain, we can somewhat mitigate this concern by permuting each replication such that the state preparation, instruction ordering, and entanglement differs while still preserving the final output. We leverage the inherent limitation of connectivity-constrained architectures to force this differentiation at transpile time, thereby weakly differentiating between replicates. With superconducting architectures, such as IBM's heavy-hex design \cite{2021/IBM-Heavy-Hex}, quantum programs must be transpiled to produce an equivalent program that is constrained to the target hardware's instruction set and connectivity. This problem is known to be NP-hard \cite{2023/Ito,2023/Wagner}, and state-of-the-art solutions use stochastic methods to perform the necessary layout and routing \cite{2024/Zou}. While normally a source of unwanted noise, we can leverage this process by transpiling all replicates together to produce subprograms that are computationally homogeneous but structurally heterogeneous.

\subsection{Quantum Approximate Optimization Algorithms}

The Quantum Approximate Optimization Algorithm (QAOA) is a method that formulates classically intractable optimization problems as Variational Quantum Circuits (VQCs) \cite{2014/Farhi,2018/Moll}. This hybrid algorithm works by evaluating a parameterized cost function encoded as a Hamiltonian to find an approximately optimal solution. A classical optimizer iteratively adjusts the parameters to minimize the Hamiltonian's latent energy to find the lowest-energy state representing an approximately optimal solution \cite{2018/Moll}. Because of its iterative approach, QAOA is highly sensitive to noise \cite{2018/Moll,2024/Nath}, making it a good candidate for our analysis.

The Maxcut problem takes an undirected, weighted graph and attempts to partition its nodes so that the weight of edges between the partitions is maximal. This problem is NP-complete \cite{1972/Karp,2009/Garey} and further APX-hard \cite{1991/Papadimitriou}, with no classical polynomial-time algorithms or approximations that are arbitrarily close to the optimal solution. This problem is well-suited to QAOA as it involves binary classification over a finite number of observable nodes. Further, it is well-structured and well-researched, providing a good basis for comparative analysis between methods.

To simplify analysis for this work, we implement the MaxCut problem using QAOA on an $N$ node graph with an arbitrary number of edges. We encode each node as a single qubit and each edge as an observable Pauli pair within the cost Hamiltonian. While there may be more qubit-efficient methods for encoding this data \cite{2026/Tene-Cohen}, our analysis focuses on profiling replication as an error-mitigation strategy rather than on the performance of the workload under an arbitrary error profile.

\section{Methodology}
\label{sec:methodology}

In this section we briefly present the methodology used to collect and compare data. We also discuss how we establish correctness for the purposes of this analysis.

\subsection{Data Collection}

\begin{figure*}
  \centering
  \subfloat[]{%
    \includegraphics[width=0.15\linewidth]{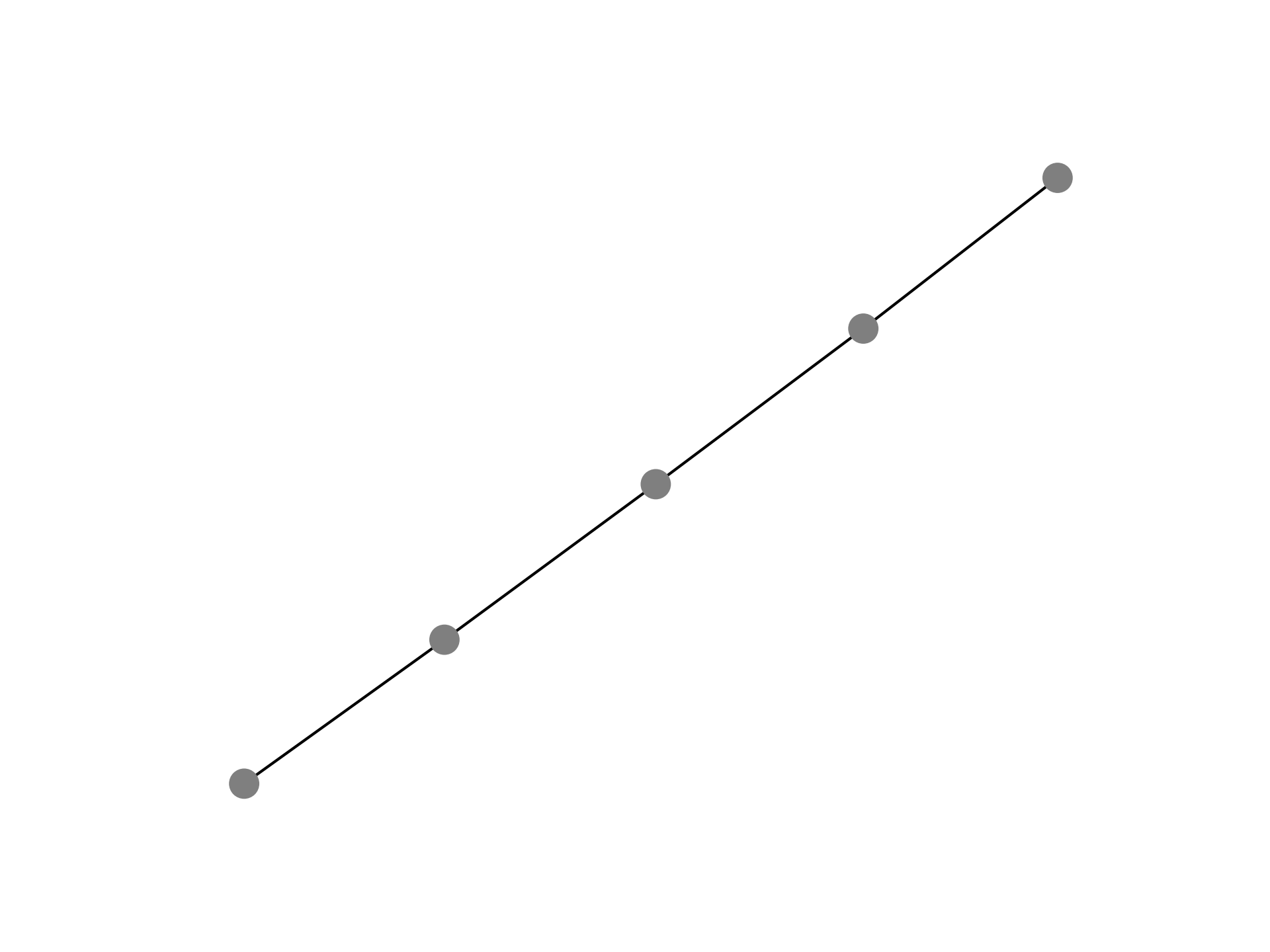}}
    \hspace{1em}
  \subfloat[]{%
    \includegraphics[width=0.15\linewidth]{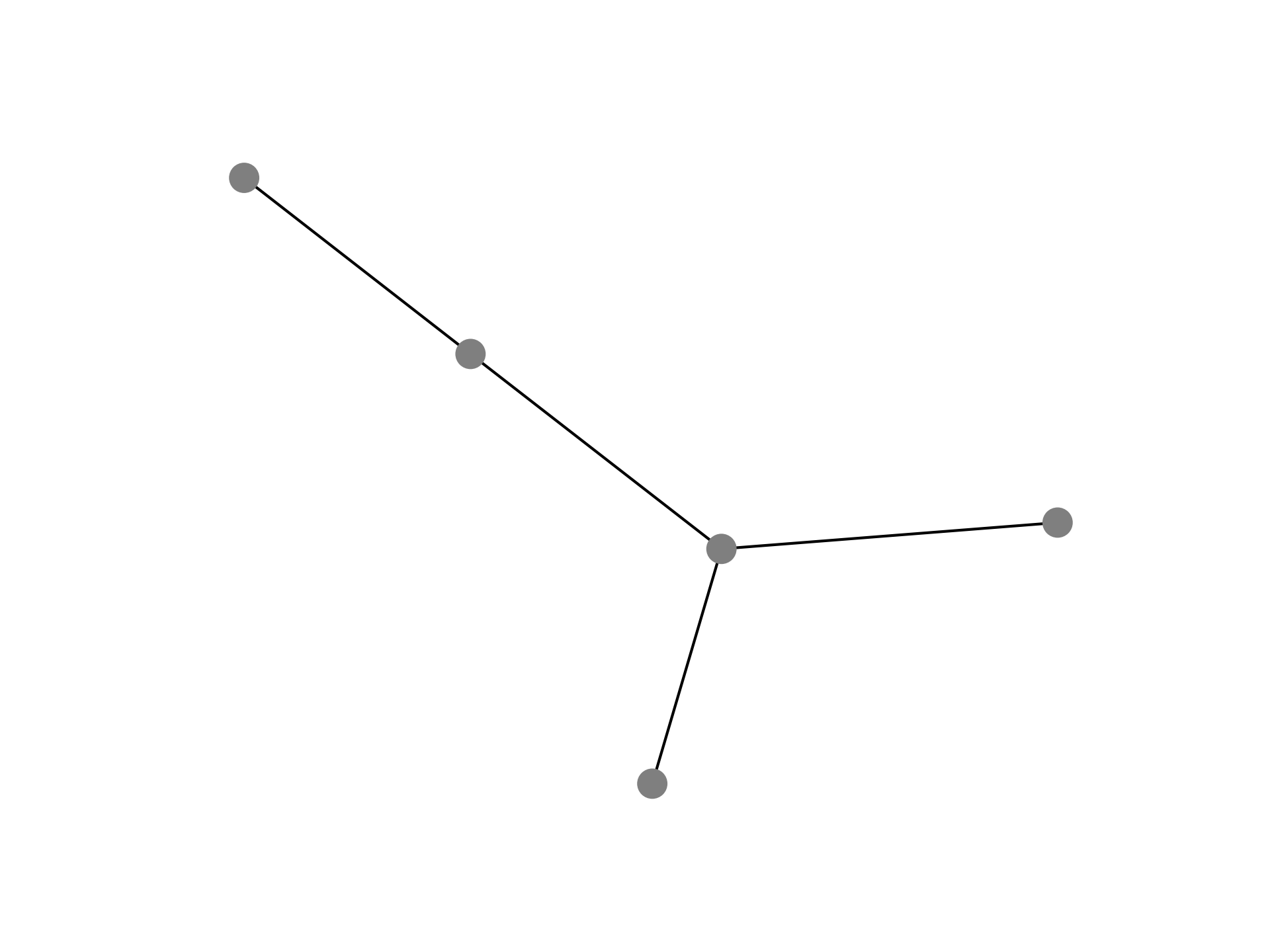}}
    \hspace{1em}
  \subfloat[]{%
    \includegraphics[width=0.15\linewidth]{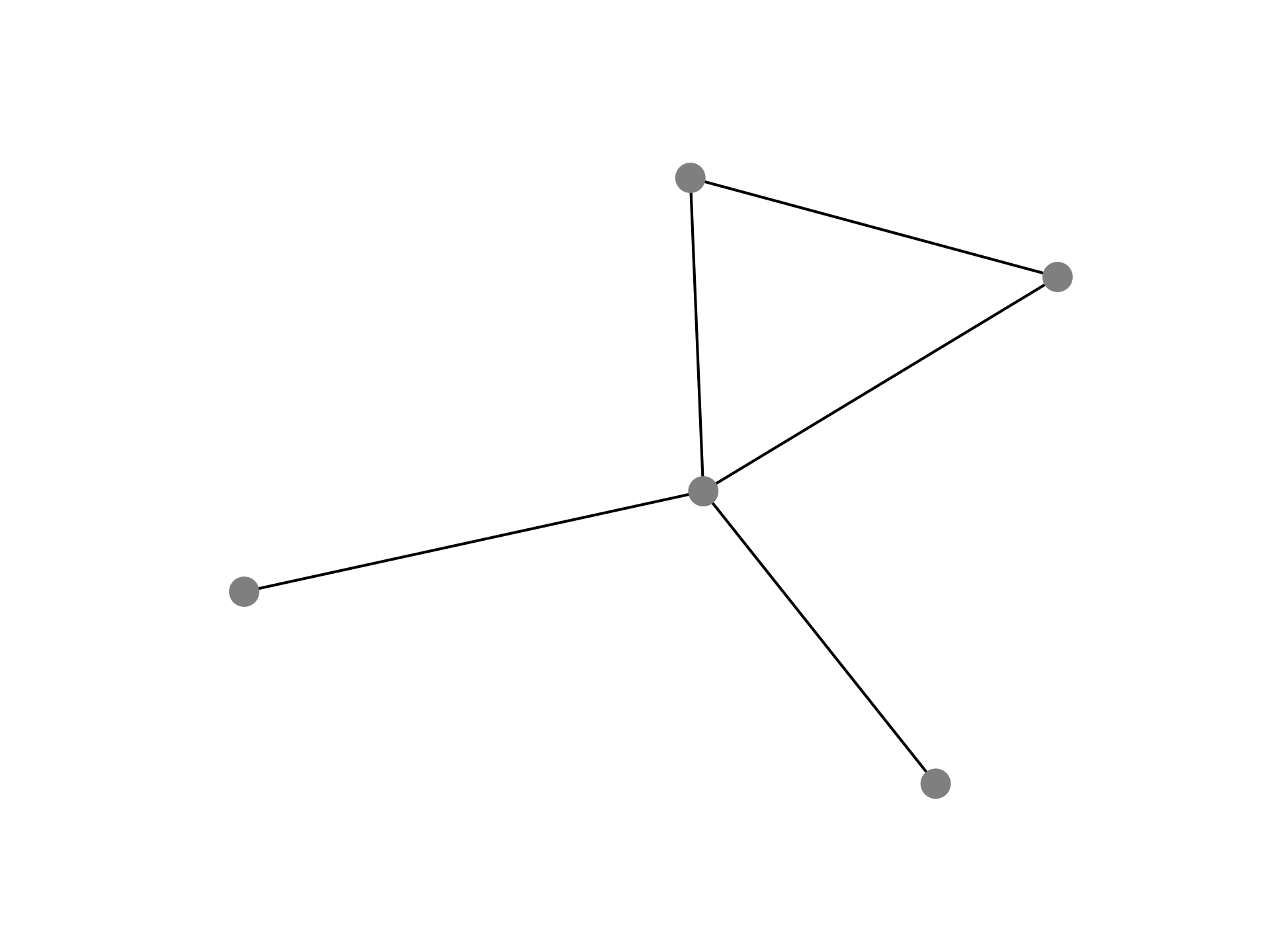}}
    \hspace{1em}
  \subfloat[]{%
    \includegraphics[width=0.15\linewidth]{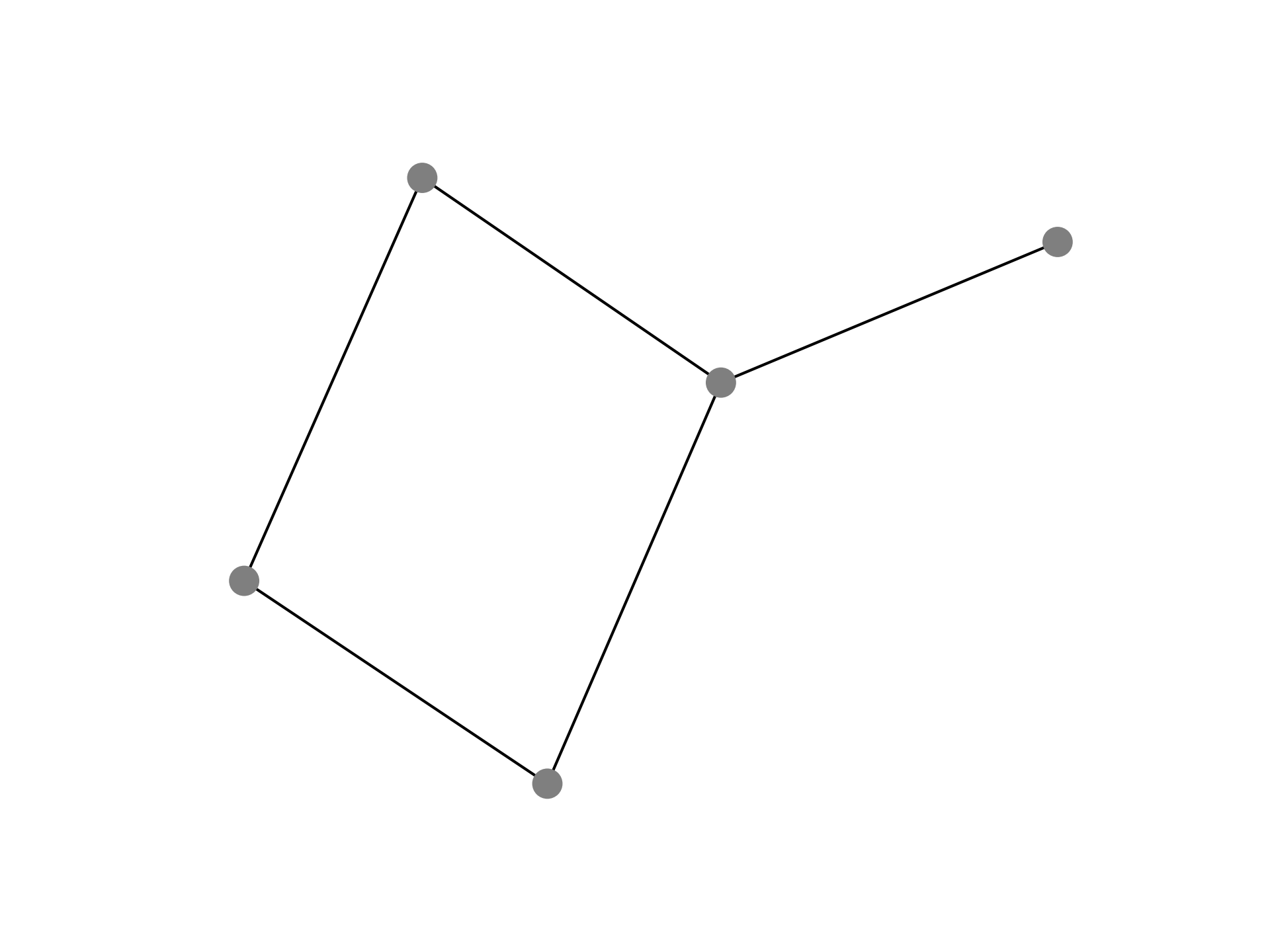}}
    \hspace{1em}
  \subfloat[]{%
    \includegraphics[width=0.15\linewidth]{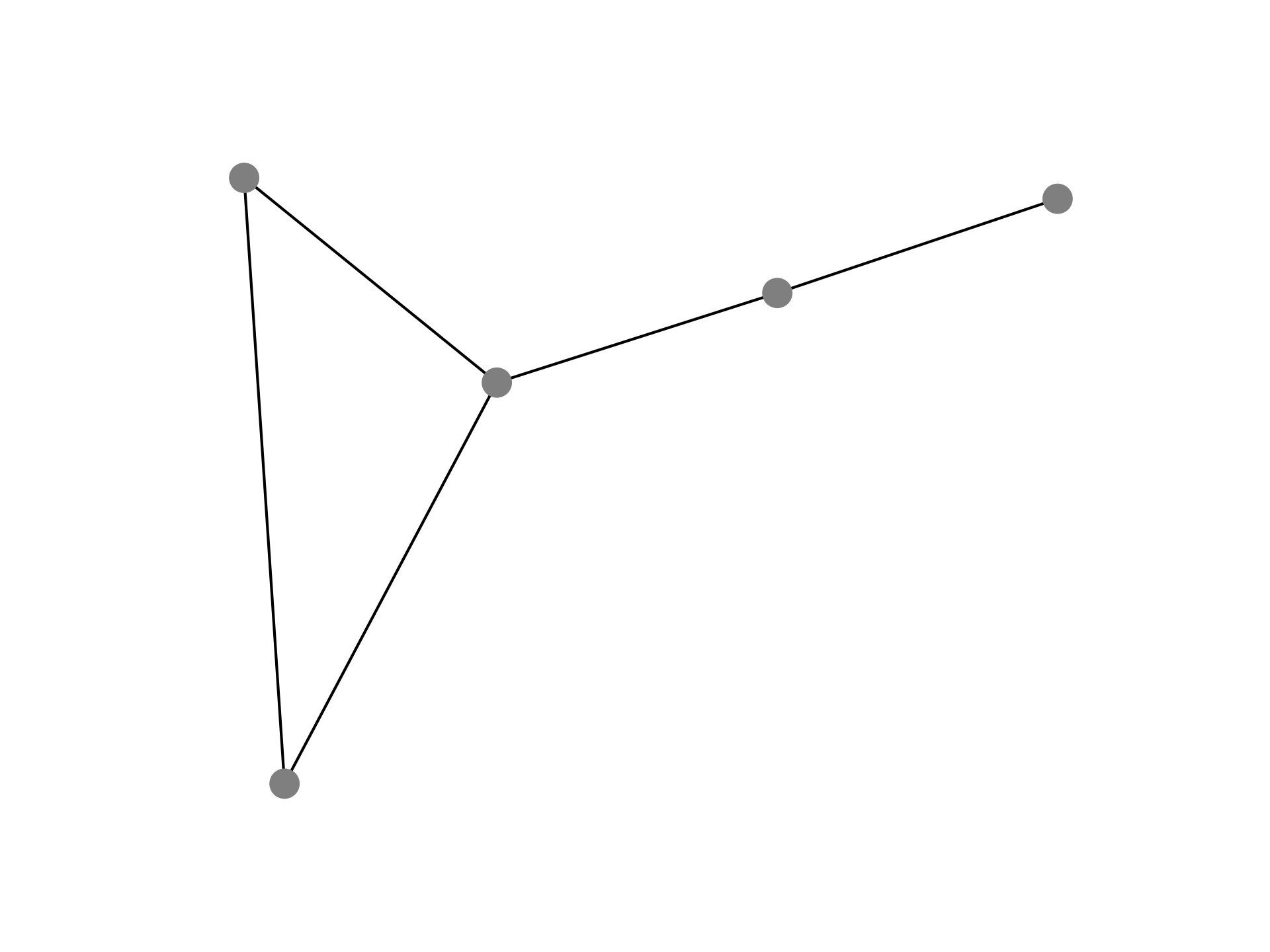}}
  \linebreak
  \subfloat[]{%
    \includegraphics[width=0.15\linewidth]{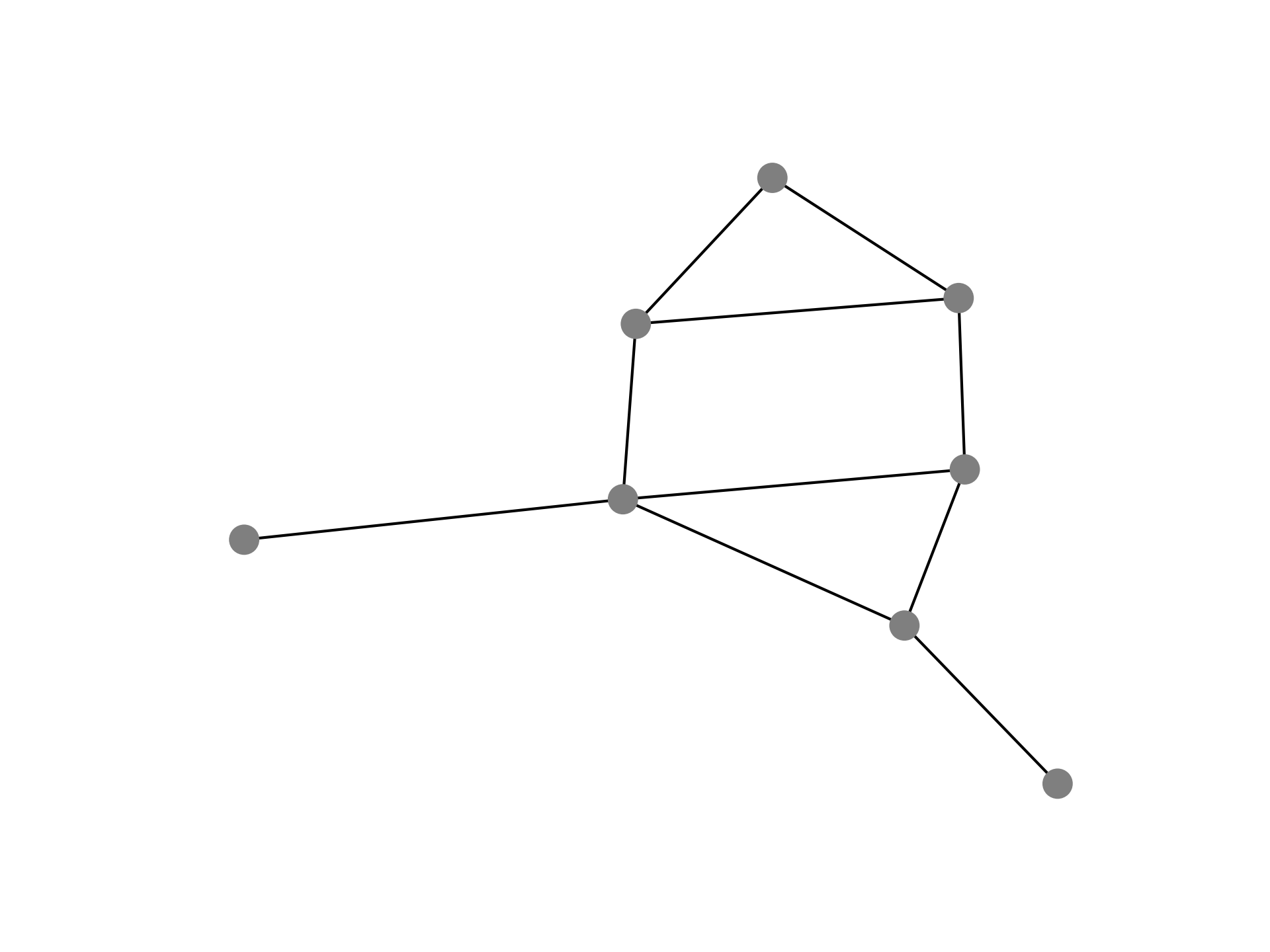}}
    \hspace{1em}
  \subfloat[]{%
    \includegraphics[width=0.15\linewidth]{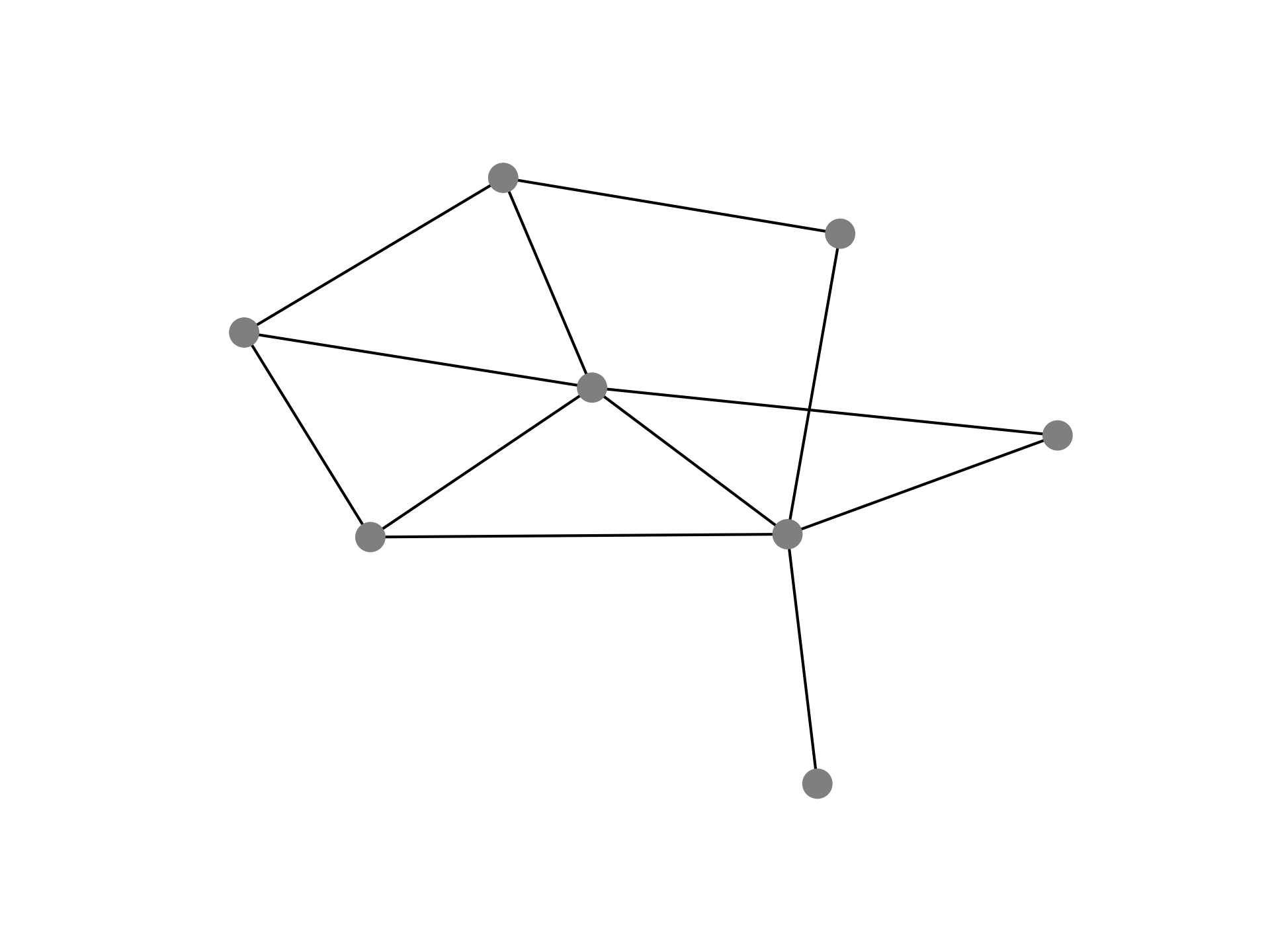}}
    \hspace{1em}
  \subfloat[]{%
    \includegraphics[width=0.15\linewidth]{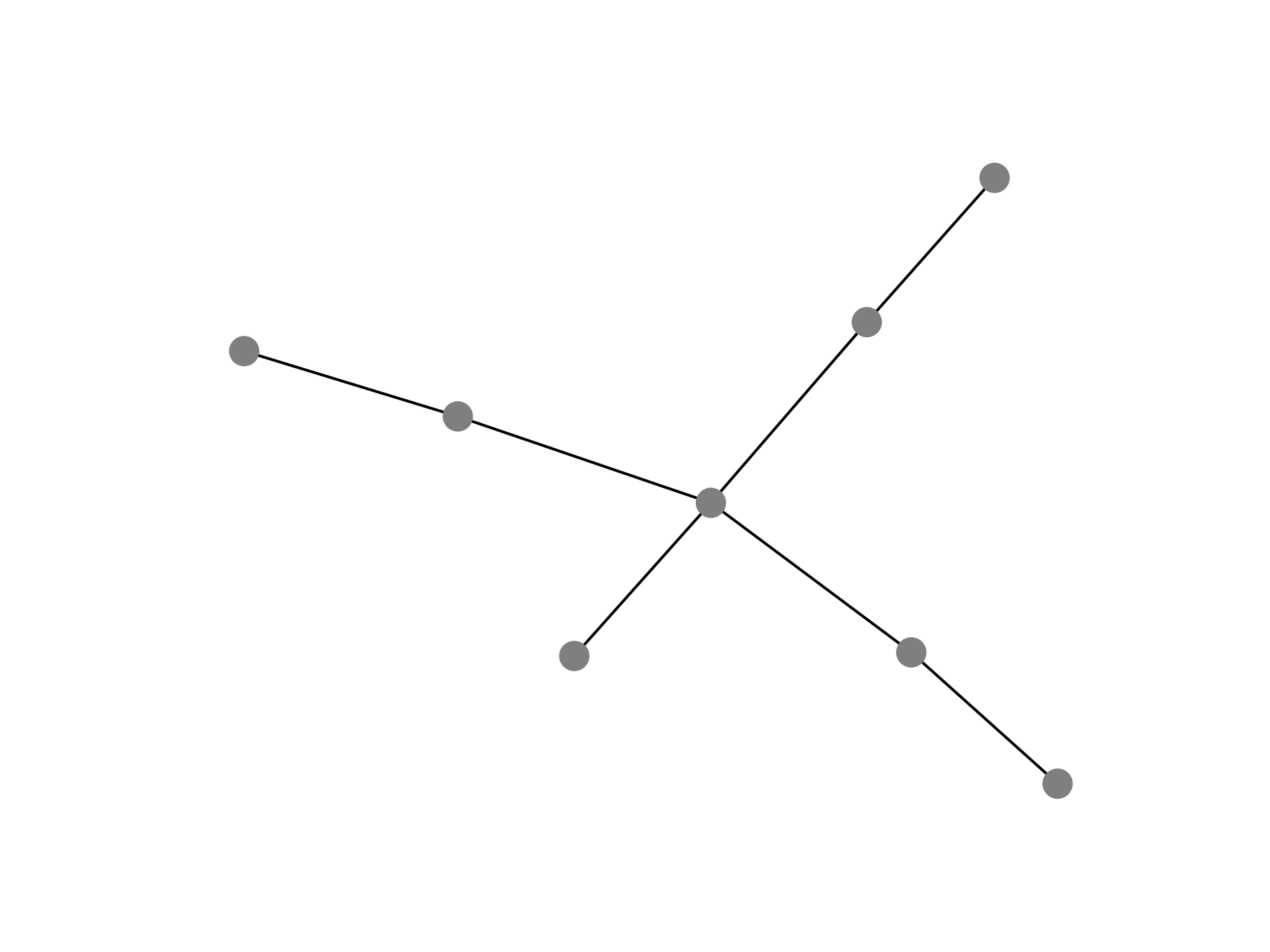}}
    \hspace{1em}
  \subfloat[]{%
    \includegraphics[width=0.15\linewidth]{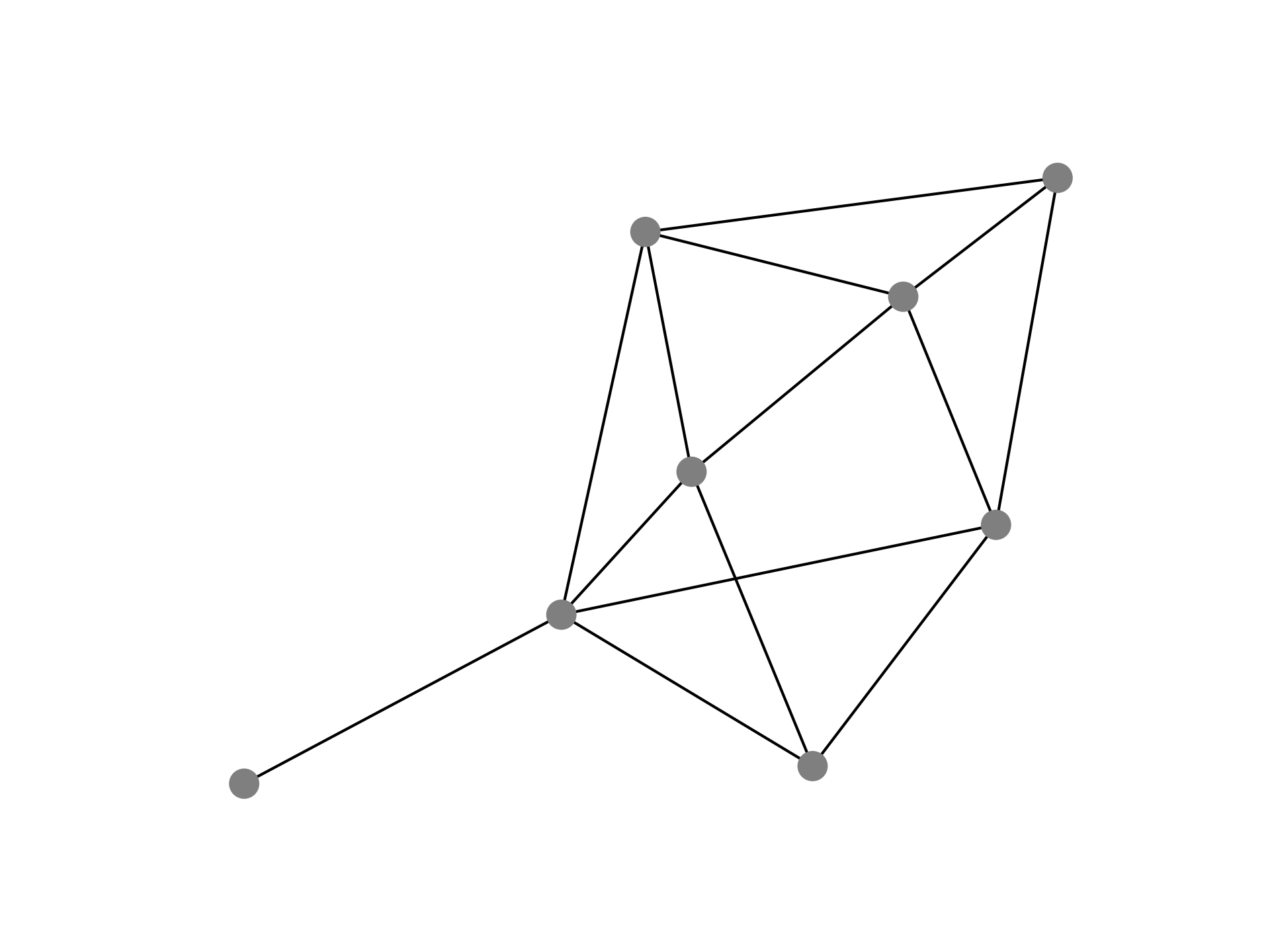}}
    \hspace{1em}
  \subfloat[]{%
    \includegraphics[width=0.15\linewidth]{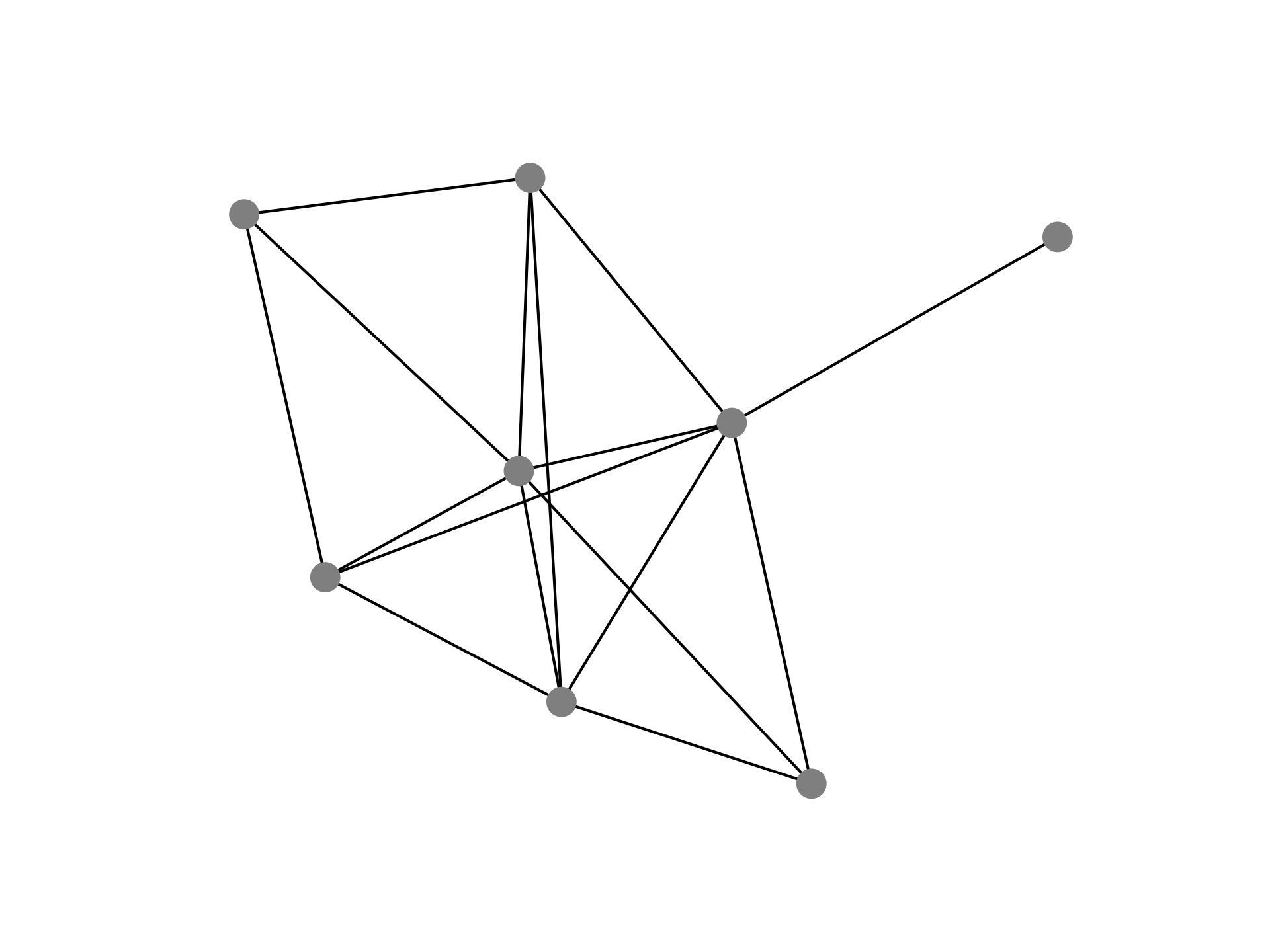}}
  \caption{The complexity of finding the maximal cut for these graphs scales with the number of nodes and edges. For $N=5$ graphs (shown as (a) through (e)), the solution is trivially computable. For $N=8$ (shown as (f) through (j)), the solution is complicated by the varying number of edges between graphs.}
  \label{fig:graphs}
\end{figure*}

We chose four sets of graphs of size $5$, $6$, $7$, and $8$, each populated with five randomly generated, undirected graphs using the Erdős-Rényi method with an edge probability of $40\%$. By allowing the number of edges to differ across graphs, as shown in \autoref{fig:graphs}, we can capture a more meaningful representation of the overall problem.

After generation, all graphs are translated into a nominal quantum form using QAOA. Nodes of the graph directly correspond to the qubits in the quantum program, while edges are encoded as sparse Pauli-pairs of interactions between qubits. To limit the total number of parameters and reduce potential circuit depth, we use only two QAOA stages during program construction.

To transpile the program, we use the Qiskit Aer simulator targeting IBM's $156$-qubit Fez platform via the \texttt{fake\_fez} provider. For consistency, we do not allow the hardware model's noise model to influence the transpiler directly. Instead, we mask the platform by using only the hardware's connectivity and basis-gate set for transpilation. This allows for later simulation to be performed using a strictly controlled noise model.

To perform replication, we use a custom transpiler pass to duplicate, transpile via the default, conventional means, and decompose the input program. We gather data for replication levels starting at $1$ (no replication) to $6$ ($6$ parallel programs). The program decomposition is necessary to extract each replicate for simulation while still allowing the routing and layout to be replicate-aware. Due to the way the Qiskit transpiler works, there is no guarantee that the act of transpilation will preserve the linear separability of each of the replicates. This quirk is an unavoidable feature of the platform, as two otherwise separate programs could become intertwined by sharing an ancilla. For cases where this happened to more than two subprograms, we repeat the transpilation pass using a new seed.

After transpilation, each subprogram is simulated in parallel. We treat each subprogram as a parallel shot, meaning that the final probability output retains the shape of the initial input program regardless of duplication. The entire process is then repeated ten times to provide an more stable average basis.

\subsection{Metrics of Analysis}

To compare results consistently, we adopt the methodology used by Tannu et al. \cite{2019/Tannu}. When establishing correctness, quantum programs executed using NISQ computers are characterized by the \textit{Probability of Successful Trial} (PST). This probability
compares the number of successful trials to the total number of trials observed as a ratio, providing good visibility into the raw fidelity of the quantum computer but poor comparability between machines and workloads.

For the Maxcut problem in particular, the deficiencies of this approach become apparent as the workload scales. As the number of qubits required to represent the input graph increases, the overall probability of getting the optimal answer decreases. To counteract this, we continue Tannu et al.'s analysis by comparing relative success via the \textit{Inference Strength} (IST). This metric, shown in \autoref{equ:ist}, is the ratio of the probability of a successful trial to the probability of the highest erroneous trial, providing a more comparable basis for the overall performance of a method across different workloads.
\vspace{0.1ex}
\begin{equation}
    IST = \frac{\text{Frequency of Successful Trials}}{\text{Highest Frequency of Erroneous Trials}}
    \label{equ:ist}
\end{equation}

\section{Evaluation}
\label{sec:evaluation}

We analyze the effect of replications ranging from $1$ to $6$ on graph sets of sizes $5$ to $8$ nodes. Each graph set contains five randomly generated graphs with unequal numbers of edges and data collected from ten trials run under the same parameters but with different translation seeds. All analysis was performed using IBM's $156$-qbuit Fez architecture under an empty noise model (referenced as \textit{noiseless}) as well as a noise model modeled after real calibration data (referenced as \textit{noisy}).

\subsection{Effects on Inference Strength}

Regardless of the noise model, replication had a stabilizing effect on the output, drawing the average inference strength across all graph sizes towards a median. This is not an improvement in the average inference strength across replicates, but rather an improvement in the overall stability of a circuit's output. These results, on their own, are not surprising, as replication effectively multiplies the number of shots a program executes at a given time, yielding more samples and, thus, a more stable average at higher levels of replication.

\begin{figure}
  \centering
  \includegraphics[width=0.95\linewidth]{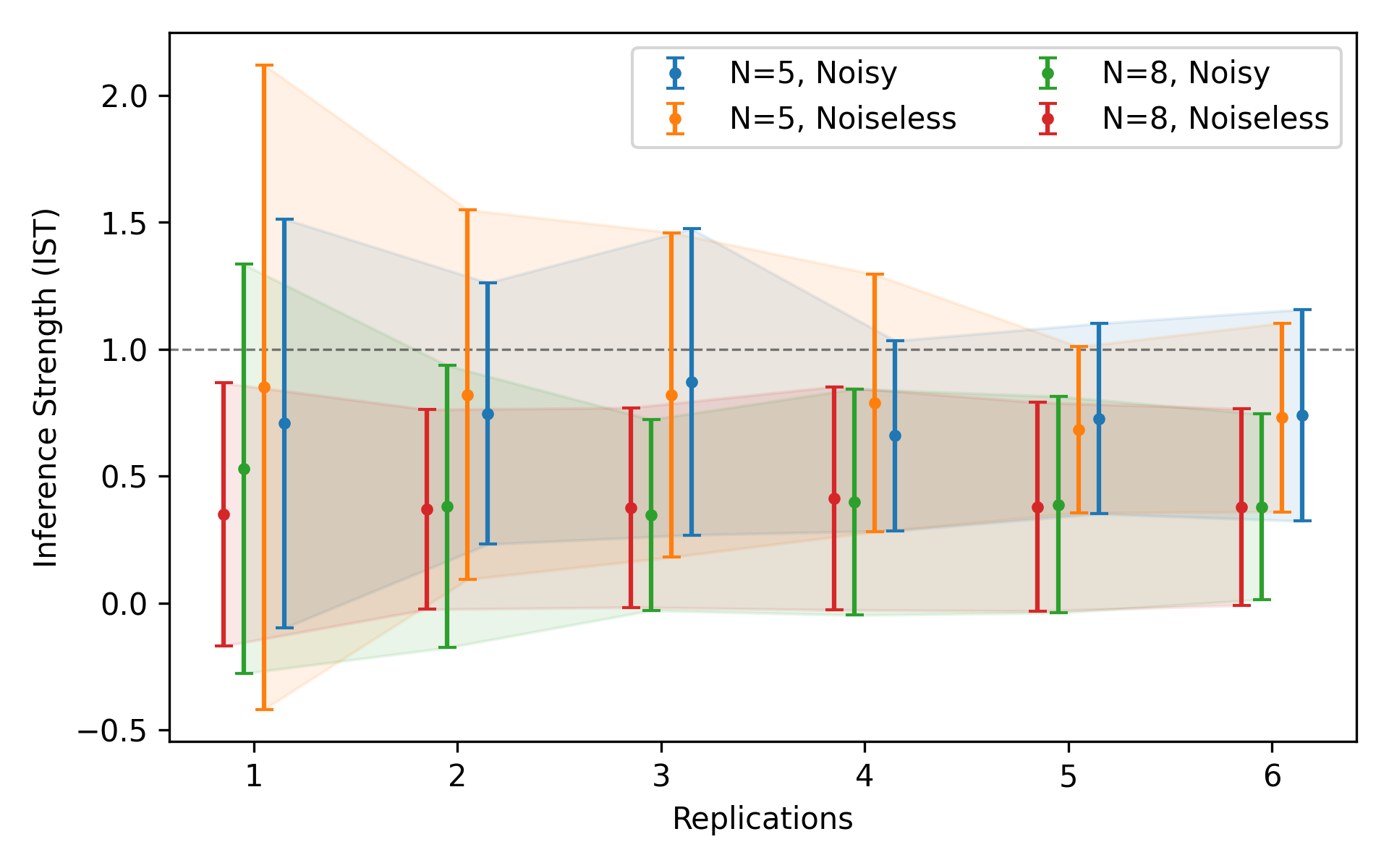}
  \caption{Average inference strength for all graphs of a specific size. Error bounds represents the standard deviation of all trials and all graphs of a specific size. A line at a strength of $1$ is drawn for visual clarity. Regardless of the noise model, inference strength improves as replication increases. This trend continues regardless of the underlying size. Both graphs also show a strong trend: the variability of the output decreases as the number of replicates increases. For noisy simulation, stability plateaus between $3$ and $4$ replications.}
  \label{fig:ist}
\end{figure}

For smaller graphs, which require fewer qubits and thus would accumulate less noise outright, the noiseless model outperforms the noisy model for replication levels $1$ and $2$. For larger graphs, the noisy model outperforms at the same replication levels. Regardless of workload complexity, both graphs show that as the number of replications increases, the overall variability in the output trends down. An unexpected trend is in the absolute average. Regardless of noise model, the average between all replication levels remains similar. For larger graphs in particular, this trend is much stronger.

\subsection{Degradation of Output Quality}

\begin{figure}
  \centering
  \subfloat[Noisy, Replicates $1$, Average of All Graphs $N=8$]{%
    \includegraphics[width=0.95\linewidth]{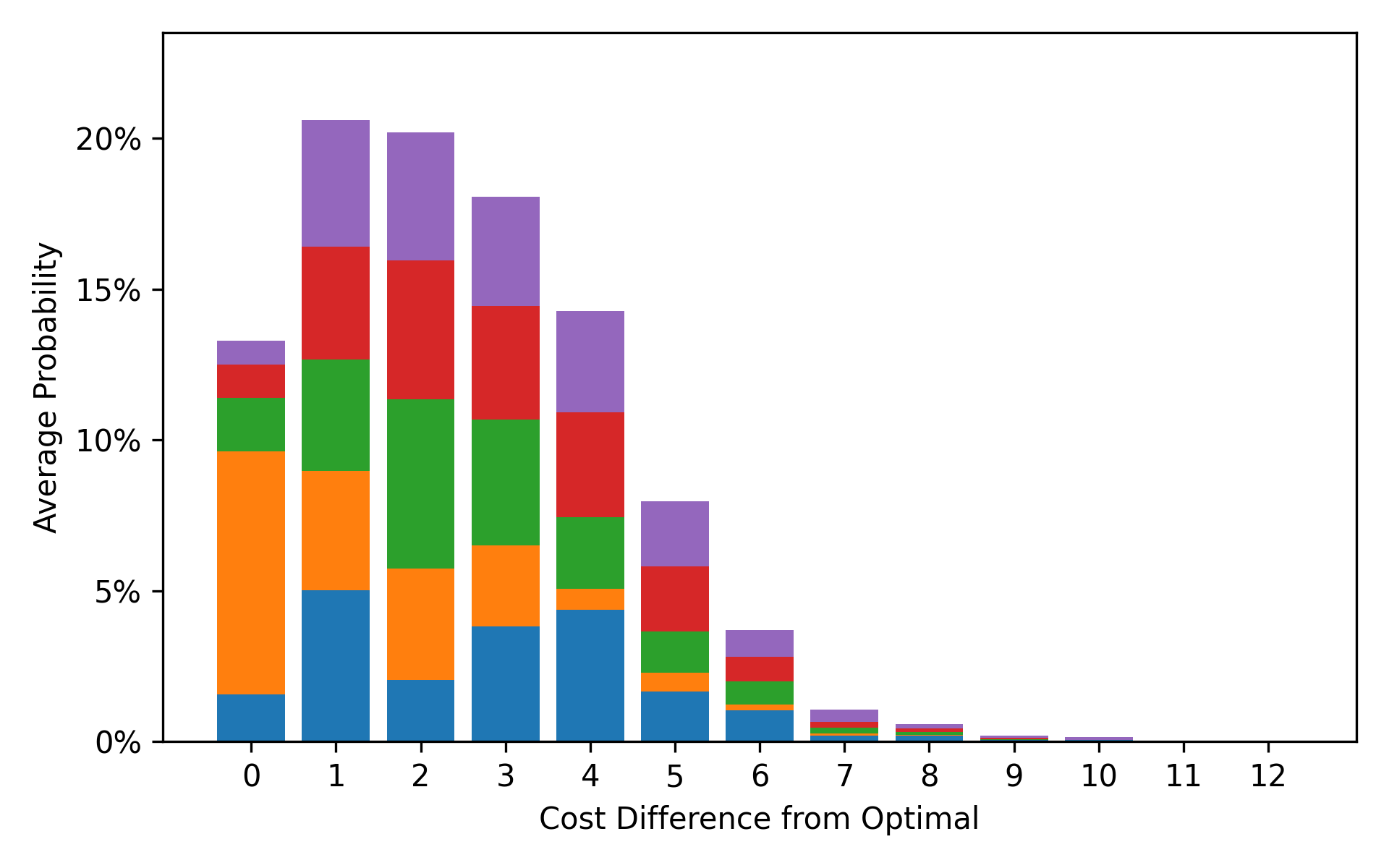}}
    \linebreak
  \subfloat[Noisy, Replicates $6$, Average of All Graphs $N=8$]{%
    \includegraphics[width=0.95\linewidth]{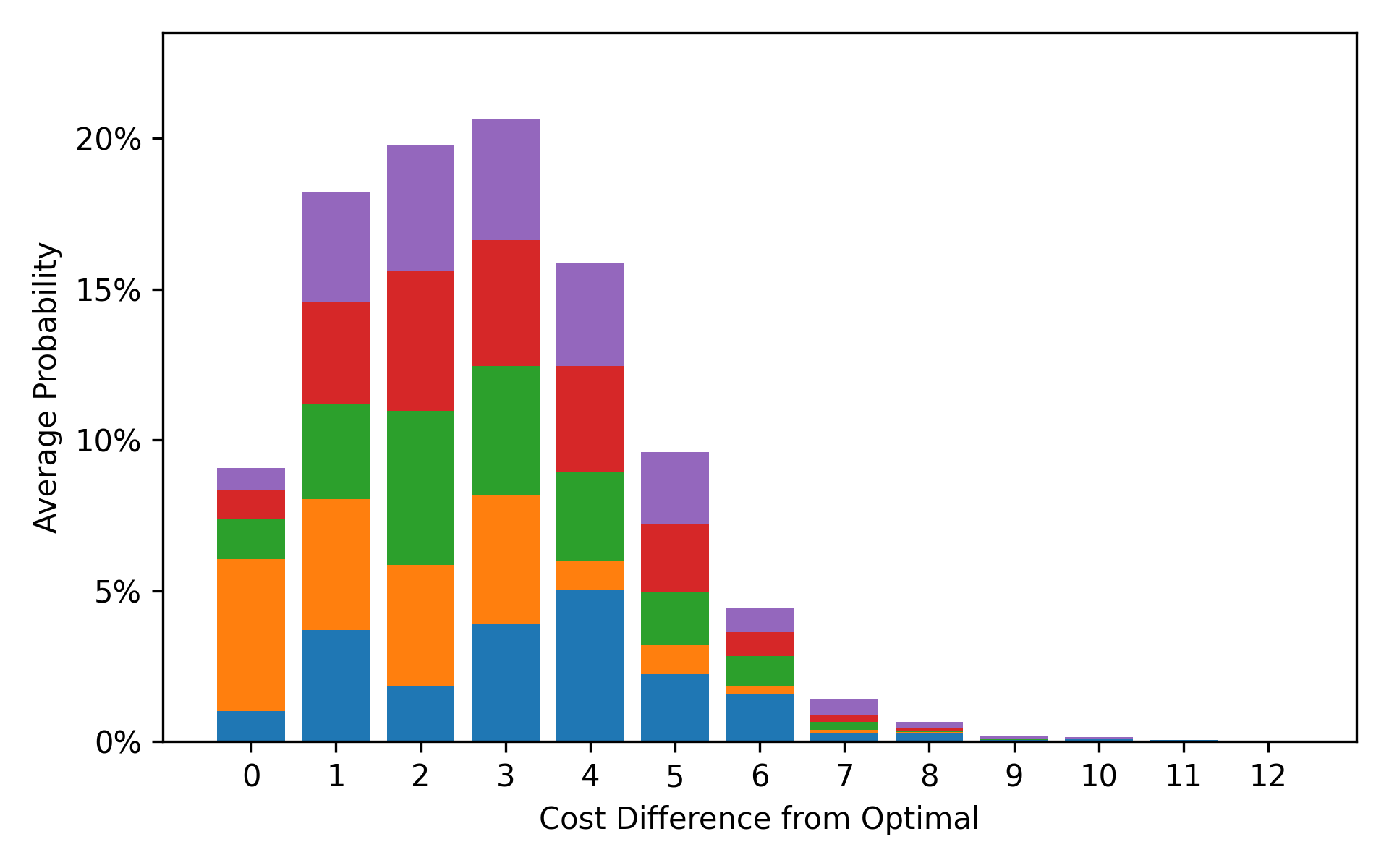}}
    \linebreak
  \subfloat{%
    \includegraphics[height=0.5cm]{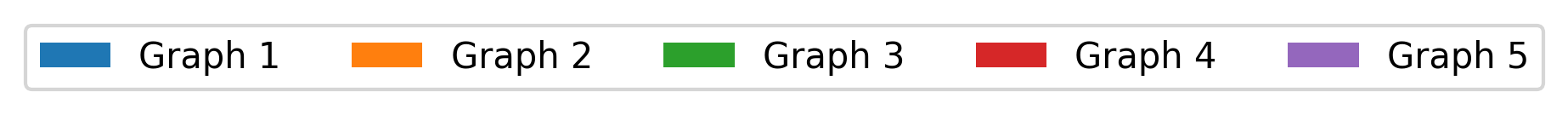}}
  \caption{Change in the relative performance of all graphs where $N=8$ under using the hardware noise profile. (a) shows the average probabilities distribution at a replication level of $1$ while (b) shows the same distribution for level $6$. The average probability drops from $13.3\%$ to $9.1\%$.}
  \label{fig:costs-changes}
\end{figure}

As would otherwise be expected, replication alone leads to a degradation of the output quality. With a greater number of quantum gates, the overall program is much more likely to accumulate error. For smaller graphs, this is less pronounced, but as is shown in \autoref{fig:costs-changes}, the larger graph set has a much more noticeable shift. Here the average probability in getting the correct result drops from $13.3\%$ to $9.1\%$, largely due to the performance of Graph 2. This is likely due to the lack of any kind of error suppression within the replicates themselves letting some workloads outperform under the noisy model for lower levels of replication.

\subsection{Limitations}
\label{sec:limitations}

\subsubsection{Limited Workload Variability}

The Maxcut problem is hard to scale in simulation, as sufficiently large graphs cannot be replicated as many times while still fitting on the target hardware. To keep this analysis simulatable, only a subset of graph sizes was utilized. Future work will implement stricter controls for graph generation and increase the number of graphs under test, especially for larger graphs. 

To further diversify our analysis, this work continues by adding other commonly used workloads, such as the Quantum Fourier Transform, the Quantum Adiabatic Algorithm, and Unstructured Search, among others. Each of these workloads has different requirements, such as entanglement and depth, which will provide a richer analysis of the kinds of workloads that replication affects.

\subsubsection{Platform Limitations}

Platform limitations prevent this analysis from fully considering the cost of replication. Transpiler algorithms are not designed with parallel or isolated program execution in mind. As a result, transpiling replicated programs attempts to compact the entire program into a single region of the hardware, despite the subprogram's linear separability. While beneficial for fully connected, highly local programs, the individual replicates end up crowded, which increases the need for SWAP's, heightens correlated error, and potentially even increases depth via shared resources. Different transpiler settings, optimization passes, routing algorithms, and other hyperparameters need to be fully tested to determine their effects on output quality.

\section{Conclusion}
\label{sec:conclusion}

In this paper, we showed that replication, while not a true means of achieving error mitigation, can be leveraged to stabilize a circuit's output regardless of the underlying workload. This stability persists even when the workload and underlying noise model are varied. While these effects alone are insufficient to operate at scale, they provide a good platform for other error-correction schemes to operate. This initial analysis indicates that further exploration into the characteristics of this behavior is warranted.

\section{Acknowledgments}

This work is supported by the National Science Foundation (CNS-2129675, CCF-2210963) and gifts from Intel.

\bibliography{references}
\end{document}